\documentclass{osa-article}

\journal{osac}

\articletype{Research Article}

\begin{document}

\title{Spectral considerations of Entangled two-photon absorption effects in Hong-Ou-Mandel interference experiments}

\author{Freiman Triana-Arango,\authormark{1,*} Gabriel Ramos-Ortiz,\authormark{1,*} and Roberto Ramírez-Alarcón\authormark{1,*}}

\address{\authormark{1}Centro de Investigaciones en Óptica AC, Apartado Postal 37150, León, Gto, México}

\email{\authormark{*}freiman@cio.mx, garamoso@cio.mx, roberto.ramirez@cio.mx} 



\begin{abstract}
Recently, different experimental methods intended to detect the entangled two-photon absorption (ETPA) phenomenon in a variety of materials have been reported. The present work explores a different approach on which the ETPA process is studied based on the changes induced in the visibility of a Hong-Ou-Mandel (HOM) interferogram. By using an organic solution of Rhodamine B as a model of nonlinear material interacting with entangled photons at $\sim800 nm$ region produced by spontaneous parametric down conversion (SPDC) Type-II, the conditions that make possible to detect changes in the visibility of a HOM  interferogram upon ETPA are investigated. We support the discussion of our results by presenting a model in which the sample is considered as a spectral filtering function which fulfills the energy conservation conditions required by ETPA, allowing to explain the experimental observations with good agreement. We believe that this work represents a new perspective to studying the ETPA interaction, by using an ultra-sensitive quantum interference technique and a detailed mathematical model of the process.
\end{abstract}

\section{\label{sec:Introduction}Introduction}
The nonlinear optical phenomenon of two-photon absorption (TPA) is of current great interest, as it finds various scientific and technological applications, such as laser scanning multiphoton microscopy, photodynamic therapy, microengraving, etc \cite{WDenk, Wei07, Aparicio-Ixta2016,Ying-Zhong-Ma,Audrey-Eshun}. Typically, TPA is achieved with nonlinear materials of large TPA cross-sections $(\delta_{c})$ through the use of pulsed lasers delivering high density of random photons. However, from recent theoretical \cite{Fei,FrankSchlawin,Bahaa-Saleh} and experimental works \cite{Dayan_exp,Goodson3}, the interest for implementing TPA has extended to the use of correlated (entangled) photons, namely entangled two-photon absorption (ETPA) process.

The advantage of ETPA compared with its counterpart, the classical TPA effect, is that in the former the total rate of absorbed photons $(R_{TPA})$ has a linear dependence on the photon excitation flux $(\phi)$, while the dependence is quadratic in the case of classical light \cite{FrankSchlawin,Perina,Dayan_teo}: $R_{TPA}=\sigma_{e}\phi+\delta_{c}\phi^{2}$, where $\sigma_{e}$ is the entangled TPA cross-section. Different authors claim there is a difference up to $\sim 32$ orders of magnitude between $\sigma_{e}$ \cite{Goodson3,Tabakaev,Goodson4} and $\delta_{c}$ \cite{Makarov08,Xu96,Sperber}, with estimated values of $\sigma_{e}$ for molecules used as nonlinear models falling within the wide range from $10^{-22}$ to $10^{-18}[cm^{2}/molecule]$ \cite{JuanVillabona1, Tabakaev, KristenM}. Although it is not completely correct compare directly these two parameters because they have different units $(\sigma_{e}[cm^{2}/molecule],~\delta_{c}[cm^{4}s/molecule])$, in principle it would be possible to achieve ETPA in the low flux regime by illuminating the sample with correlated photon pairs produced by spontaneous parametric down conversion (SPDC) \cite{SPDC_Adel}. Nevertheless, there is a recent debate about the actual magnitude of $\sigma_{e}$ or even if the ETPA effects has been really detected experimentally \cite{KristenM,Samuel_Alfred}. 

Diverse experimental configurations have been employed to study the ETPA activity, based on measuring the transmittance \cite{Goodson3,Goodson4,Goodson1,JuanVillabona1,Samuel_Alfred} or fluorescence exhibited by the sample upon excitation with SPDC photon pairs \cite{Dayan_exp,Goodson2,KristenM,TiemoLandes}. For the purposes of the present work, we highlight the case where the difference in transmittance between a solution of the molecule under test and the solvent alone is recorded as a function of the excitation intensity, where the temporal delay ($\delta t$) between the entangled photons is set to zero. In some of these works, the zero delay configuration is assured by implementing a HOM interferometer \cite{Samuel_Alfred} before the photons interact with the sample. It is recognized  that in these experimental attempts to detect ETPA the results can be biased by artifacts\cite{KristenM,Landes:21}. The artifacts emulating the ETPA process are: optical losses induced by linear absorption, scattering, molecular aggregation, hot band absorption and others. The existence of these artifacts\cite{KristenM,AlexanderM,AJAMI2015524} have fed the controversy about if the ETPA has been or not demonstrated unambiguously in experiments. In this context it is worth to further investigate ETPA, imposing experimental conditions, including the discrimination of artifacts, that contribute to discard or confirm the possibility of observing the effect. 

Recently, the HOM interference based in a Mach-Zehnder interferometer was used to analyze changes in entangled photon pairs, where the time delay shifting induced by a sample in the HOM interferogram (HOM dip) is utilized as a parameter to perform linear spectroscopy studies \cite{Audrey_Goodson}. Likewise, it has been shown that the features of the HOM dip can provide a signature of the linear absorption spectrum of a resonant sample with the two photon spectrum\cite{Kalashnikov2017,Dorfman2021}. Although the HOM dip has already been used to explore ETPA experiments\cite{AlexanderMikhaylov2020}, the visibility change in a HOM dip has not received attention as a parameter aimed to analyze  the ETPA phenomenon. So, with the motivation of detecting nonlinear effects, new approaches in the use of HOM interference deserves detailed examination, taking advantage of its sensibility and robustness to discriminate photon states. In this work, we present a detailed study of the effects produced on the visibility of a HOM dip by a nonlinear sample illuminated with entangled photons. Experimentally, the pairs of entangled photons interact with the sample under test before they are launched into the HOM interferometer. This approach allows to explore directly the changes in the entangled photon states, while by discriminating the linear optical losses the conditions and possibilities to detect ETPA are established. This is the first time, to the best of our knowledge, that the visibility of a HOM dip is fully investigated intending to detect ETPA.

\section{\label{sec:Mathematical-model}Theory}
\subsection{\label{sec:Mathematical-model}Mathematical model}
 For the discussion of the experimental results obtained from HOM interferograms presented in the next section, here we propose a model based on the two-photon transitions induced in a nonlinear sample by down-converted (entangled) photons. In our case, the HOM interference is produced after the interaction of down-converted photons with the sample. The model assess the spectral-temporal conditions occurring during the ETPA effect, and therefore, allows to realize in which cases the sample produces a measurable effect over the joint spectral intensity (JSI) function of the interfering photons. Since the features of the HOM dip are directly determined by the JSI function, any effect of the ETPA over such function might be observed as a modification in the visibility of the HOM dip.
	
The state generated by a Type-II SPDC process can be written as: \cite{SPDC_Adel,SPDC_Adel_2,SPDC_Sergienko}
\begin{equation}
	\lvert \psi(t)\rangle=\int_{0}^{\infty}\int_{0}^{\infty}d\omega_{s}d\omega_{i}\zeta(\omega_{s},\omega_{i})\lvert \omega_{s} \rangle_{s}\lvert \omega_{i} \rangle_{i},
	\label{estado2}
\end{equation}	
\noindent where $\lvert\ \omega_{j} \rangle$ represents a single photon with frequency $\omega_j$ for the signal (s) or idler (i) mode, with $j=s,i$. The function $\zeta(\omega_{s},\omega_{i})$ represents the joint spectral amplitude function (JSA) and $I(\omega_{s},\omega_{i})=\left\|\zeta(\omega_{s},\omega_{i})\right\|^{2}$ is the JSI function. The JSA function contains all the relevant information of the photon pair quantum state, such that the ETPA process would produce a spectral filtering of photons, expressed as: 
\begin{equation}
    \zeta(\omega_{s},\omega_{i})=\alpha(\omega_{s},\omega_{i})\phi(\omega_{s},\omega_{i}) f(\omega_{s})f(\omega_{i}) h(\omega_{s},\omega_{i}),
    \label{JSA}
\end{equation}
\noindent where the pump envelope function $\alpha(\omega_{s},\omega_{i})$ that leads to the generation of down-converted photons, and assures the energy conservation condition ($\omega_{s}+\omega_{i}=\omega_{p}$), is given by:
\begin{equation}
    \alpha(\omega_{s},\omega_{i})=e^{-\frac{\left(\omega_{s}+\omega_{i}-\omega_{p}\right)^{2}}{2\Delta\omega_{p}^{2}}},
	\label{PUMP}
\end{equation}		
\noindent with $\omega_{p}$ and $\Delta\omega_{p}$ the central frequency and the bandwidth of the pump spectrum, respectively. In the JSA the phase-matching function $\phi(\omega_{s},\omega_{i})$ denoting the conservation of linear momentum ($\Vec{k_{s}}+\Vec{k_{i}}=\Vec{k_{p}}$) is represented by	
\begin{equation}
	\phi(\nu_{s},\nu_{i})=\frac{\sin{[\left(\tau_{s}\nu_{s}+\tau_{i}\nu_{i}\right)/2]}}{\left(\tau_{s}\nu_{s}+\tau_{i}\nu_{i}\right)/2}\simeq e^{-\frac{\gamma}{4}\left(\tau_{s}\nu_{s}+\tau_{i}\nu_{i}\right)^{2}}, 
	\label{PM}
\end{equation}		
\noindent being $\nu_{s}=\omega_{s}-\omega_{0}$, $\nu_{i}=\omega_{i}-\omega_{0}$, and  $\omega_{0}=\frac{1}{2}\omega_p$ the central frequency of the down-converted photon wavepackets while the parameter $\gamma\simeq 0.15065$ results from the sinc to gaussian function FWHM approximation. The constants $\tau_{j}$ take into account the group velocity of the photons wave-packets, defined in the Supporting Information. Equation \ref{PM} shows that the phase-matching function is not symmetric in its frequency arguments since $(\tau_{s}\neq\tau_{i})$, as a consequence of the different refractive index experimented by each photon wavepacket inside the nonlinear uniaxial crystal \cite{SPDC_Grice}.

In the JSA function given in Eq. \ref{JSA}, the terms $f(\omega_{j})$ account for the effect of the bandpass filter ($F_1$) used (see experimental section) to control the bandwidth of the down-converted photons, which has an intensity profile that can be obtained from the manufacturer datasheet, and modeled as:
\begin{equation}
	F(\omega_{j})=\left\| f(\omega_{j}) \right\|^{2}=e^{-\left[\frac{(\omega_{j}-\omega_{F})^{2}}{2\Delta\omega_{F}^{2}}\right]},
	\label{FILTER}
\end{equation}
\noindent being $\omega_{F}$ and $\Delta\omega_{F}$ the central frequency and the bandwidth of the filter, respectively.

In our analysis the ETPA effects induced by the sample on the JSA function are conveniently modeled by a filtering function $h(\omega_{s},\omega_{i})$ which fulfills the conservation of photon energy, that is, the frequency sum of the two down-converted photons must be equal to some frequency transition of the material $\omega_{H}$, namely, $\omega_{s}+\omega_{i}=\omega_{H}$. In the frequency domain, the sample transfer function can be modeled as a kind of "notch" filter \cite{Kalashnikov2017} with a Gaussian profile and bandwidth $\Delta\omega_{H}$, proposed as: 
\begin{equation}
	H(\omega_{s},\omega_{i})=\left\| h(\omega_{s},\omega_{i}) \right\|^{2}=1-\eta e^{-\frac{\left(\omega_{s}+\omega_{i}-\omega_{H}\right)^{2}}{2\Delta\omega_{H}^{2}}}-\eta',
	\label{SAMPLE}
\end{equation}		
\noindent where the parameter $\eta$ ($0\leq \eta \leq 1-\eta'$) plays a fundamental role in the modeling, since it quantifies the efficiency of the ETPA process, which depends on the sample properties (nonlinearities of the molecules conforming the sample, molecular concentration, etc.), while the parameter $\eta'$ stand for the linear optical losses independent of frequency. 

All the functions described above (pump, phase-matching, filter and sample transfer) are plugged into Eq. S1 (see Supporting Information) to simulate HOM dips obtained under different experimental conditions. We denote the HOM dip for a sample under study as $HOM_{sam}$. The $HOM_{sam}$ obtained experimentally, and reproduced theoretically by Eq. S1, can be conveniently expressed as \cite{SPDC_Grice,SPDC_Timothy,AymanF}:
\begin{equation}
	CC(\delta t)=CC_{max}-\kappa_{sam} e^{-\frac{\delta t^{2}}{2\Delta t^{2}}} ,
    \label{HOM_interferogram}
\end{equation}
\noindent where $\Delta t$ is the temporal width of the HOM dip while $\kappa_{sam}$ = $(CC_{max}-CC_{min})_{sam}$, being $CC_{max}$ and $CC_{min}$ the maximum and minimum coincidence counts ($CC$) registered experimentally. It is clear that $CC_{max}$ is registered when the time delay $\delta t$ between the entangled photons is long. On the contrary, as $\delta t$ tend to zero, $CC$ reaches its minimum possible level ($CC_{min}$), which will be zero only if JSI function is fully symmetric. The symmetry of the JSI is defined with respect to the anti-diagonal in the $(\omega_{s},~\omega_{i})$ 2D map. If  any of the functions in Eq. \ref{JSA} introduce asymmetries, $CC_{min}$ will be different from zero; as larger the asymmetry, $CC_{min}$ differ more significantly from zero. Equation \ref{HOM_interferogram} also stands for the HOM dips obtained from a solvent ($HOM_{sol}$) or from photons propagating in free space ($HOM_{ref}$). where again $CC_{max}$ ($CC_{min}$) is the maximum (minimum) level of $CC$, respectively, and $\kappa_{sol,~ref}$ = $(CC_{max}-CC_{min})_{sol,~ref}$.

The visibility of the HOM dip is given by \cite{Hong_Ou_Mandel1}:		
\begin{equation}
	V_{i}=\left(\frac{CC_{max}-CC_{min}}{CC_{max}+CC_{min}}\right)_{i}
	\label{VIS},
\end{equation}
\noindent where $i=sam,~sol,~ref$, such that its value will be dependent on the symmetry of the quantum state. The figure of merit $V_{sam}/V_{sol}$, weights how much the sample visibility ($V_{sam}$) changes from the solvent visibility ($V_{sol}$). 
Following with our model description, an expression accounting for ETPA rate ($R_{TPA}$) as a function of the sample and solvent visibilities is established (a complete deduction is found in the Supporting Information): 
\begin{equation}
    R_{TPA}=\kappa_{sol}\left(1-\frac{V_{sam}}{V_{sol}}\right),
    \label{RateETPA}
\end{equation}
\noindent In this expression $\kappa_{sol}$ accounts the amount of photons available to be absorbed by the sample (transmitted photons by solvent) per unit of time. Let us check the boundary cases of Eq. \ref{RateETPA}: when a sample of null TPA is tested, $V_{sam}=V_{sol}$, then $R_{TPA}=0$; in contrast, for the ideal case of a sample able to absorb all the collection of photons by TPA, $V_{sam}=0$, and then $R_{TPA}=\kappa_{sol}$. In this context, the possibility of detecting ETPA is associated to the ability to discriminate real differences between $V_{sam}$ and $V_{sol}$. It is important to mention that Eq. \ref{RateETPA} is independent of any optical loss other than TPA (Fresnell losses, scattering, linear absorption), since none of these optical losses change the visibility of the HOM dip and only reduce the maximum level of $CC$ as it will be discussed latter. 
\subsection{\label{sec:Mathematical-model}Simulations}
\begin{figure*}[!h]
	\centering
	\includegraphics[width=\textwidth]{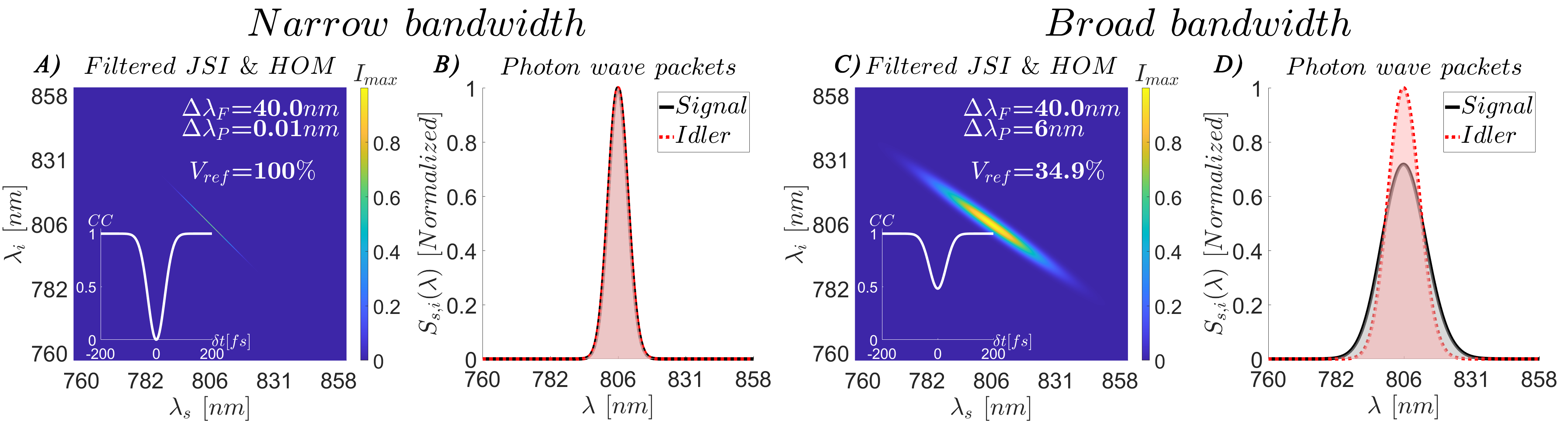}
	\caption{Theorical simulations of JSI, $HOM_{ref}$ and the singled photon wave-packets $S_{s,i}(\lambda)$ for narrow (panels $A$) and $B$)) and broad (panels $C$) and $D$)) pump bandwidths. The narrow pump bandwidth is selected to be $\Delta\lambda_{p}=0.01nm$ (FWHM) centered at $\lambda_{p}=403nm$. The broad pump bandwidth $\Delta\lambda_{p}=6nm$ (FWHM) is typical in the second harmonic generation (SHG) at $403nm$ from  femtosecond lasers with 80fs pulse-width (FWHM), emitting within the 806nm region.}
	\label{JSI_HOM_FILTER}  
\end{figure*}

The experimental HOM dip features ascribed to Eq. \ref{HOM_interferogram} result from the light-sample interaction. In its equivalent form (Eq. S1), such features are determined by sample parameters ($\omega_{H},~\Delta\omega_{H},~\eta,~\eta'$) and the entangled-photon parameters ($\phi(\omega_{s},\omega_{i})$, $\omega_{p}$, $\Delta\omega_{p}$, $\omega_{F}$, $\Delta\omega_{F}$). We first present simulations of the $HOM_{ref}$ (no sample) and the corresponding JSI. The results are plotted using the equivalent wavelengths of frequencies. We consider the case where $\omega_{F}$ corresponds to $\lambda_{F}=806nm$ for filtered down-converted photons with a bandpass filter of $\Delta\lambda_{F}=40nm$ centered at that wavelength. The phase-matching features are imposed by Eq. \ref{PM} through the parameters of the BBO Type II crystal used in our experiment, extracted directly from the manufacturer datasheet (Sellmeier Coefficients and Phase-Matching angle). Two different conditions of pump are simulated by assuming either a narrow or broad bandwidth, which can be achieved experimentally through CW and femtosecond pulsed lasers,  with bandwidths (FWHM) of $\Delta\lambda_{p}\sim 0.01nm$ and $\Delta\lambda_{p}\sim 6nm$, respectively, in the $\lambda_{p}=403nm$ region. Figure \ref{JSI_HOM_FILTER} shows the results. 

The simulations show that when the pump has a narrow bandwidth (Figure \ref{JSI_HOM_FILTER}$A$), a perfect symmetry of the JSI is obtained to produce a $V_{ref}$ of $100\%$. This maximum visibility is due to identical single photon wave-packets\cite{SPDC_Grice} (Figure \ref{JSI_HOM_FILTER}$B$), where $S_{s,i}(\lambda)$ are calculated by the integration of the JSI for each photon wavelength ($\lambda_{s},~\lambda_{i}$). The value of $V_{ref}$ diminishes below $50\%$ in the case of broad pump bandwidth (Figure \ref{JSI_HOM_FILTER}$C$) because the JSI symmetry and equivalence of the single photon wave-packets (Figure \ref{JSI_HOM_FILTER}$D$) decreased. In other words, larger pump bandwidth involves more frequencies in the SPDC process, which  essentially grows the proportion of distinguishable photons reducing the HOM interference effect. 

The model is now used for evaluating changes in the HOM dip visibility upon the interaction of the entangled photons with a sample considering the general cases of resonance ($\delta H=0$, $\lambda_{p}=403nm$ and $\lambda_{H}=403nm$) and non-resonance ($\delta H\sim 50 nm \gg 0$, $\lambda_{p}=403nm$ and $\lambda_{H}=453nm$), where $\delta H=|\lambda_{p}-\lambda_{H}|$ is the detuning wavelength between the sample and the entangled photons. Computations were performed as a function of two important sample parameters, the bandwidth of sample absorption ($\Delta\lambda_{H}$) and the strength of such absorption ($\eta$), assuming for the moment negligible linear losses ($\eta'$=0); the case where $\eta'$ > 0 is discussed later. Under these assumptions, we did not find any significant modification of the HOM dip (visibility or $CC$) of the sample with respect to the reference obtaining $V_{sam}/V_{sol}=1$ for the non-resonant case, and for the entire range of sample bandwidths even considering a sample with a large $\eta$; this prediction of negligible nonlinear effect under narrow pump bandwidth contrasts with the result previously reported \cite{Kalashnikov2017} for linear absorption spectroscopy studies using entangled photons. 

\begin{figure*}[!h]
	\centering
	\includegraphics[width=\textwidth]{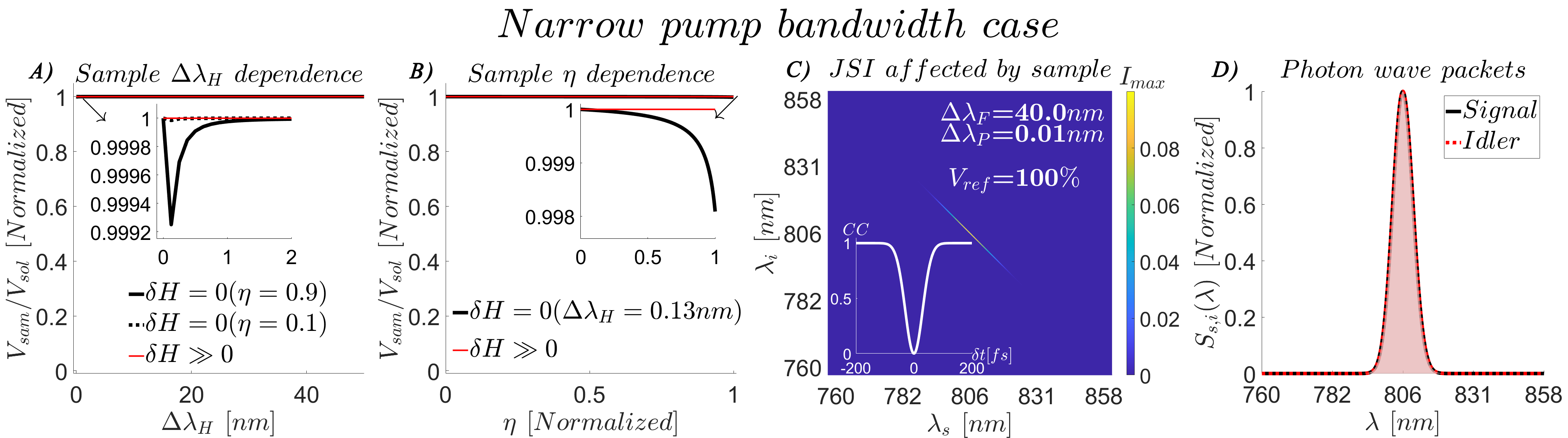}
	\caption{ETPA resulting from the sample interaction with entangled photons of narrow bandwidth. $A$) $V_{sam}/V_{sol}$ as a function of sample bandwidth, $B$) $V_{sam}/V_{sol}$ as a function of $\eta$ parameter, $C$) JSI affected by sample, $D$) single photon wave-packets. } 
	\label{CW_PUMP_CASE}
\end{figure*}

The results when a sample is tested under pump of narrow $\Delta\lambda_{p}$ are presented in the Figure \ref{CW_PUMP_CASE}.  
In panels $A$ and $B$ the simulated curves for the ratio $V_{sam}/V_{sol}$ are denoted as black and red lines for resonant and non-resonant cases, respectively. The $V_{sam}/V_{sol}$ as a function of the the sample bandwidth is shown in Figure \ref{CW_PUMP_CASE}$A$, computed for a hypothetical samples bearing a large nonlinearity ($\eta=0.9$) (continuous line) and small nonlinearity ($\eta=0.1$) (dashed line). Notice that for a large $\eta$ a minimum of $V_{sam}/V_{sol}$ is predicted at $\Delta\lambda_{H}\approx0.13nm$, but this change is less than $0.1\%$ (see inset of the figure), which  according to Eq. \ref{RateETPA} means that $0.1\%$ of the photons that hit the sample are absorbed by the sample through ETPA. Figure \ref{CW_PUMP_CASE}$B$ shows the $V_{sam}/V_{sol}$ (black line) as a function of $\eta$ for a sample bandwidth of $\Delta\lambda_{H}=0.13nm$. The ratio $V_{sam}/V_{sol}$ does not exhibit a variation larger than $0.2\%$, as it can be seen in the inset of the Figure \ref{CW_PUMP_CASE}$B$; the fact that the HOM dip remains unaltered when the photons interact with the sample means that their states do not change. Finally, the Figures \ref{CW_PUMP_CASE}$C$,$D$ show the photons states through the JSI and the singled photon wave-packets upon interaction with the sample, assuming $\eta=0.9$  and the optimum value of $\Delta\lambda_{H}=0.13nm$. It is observed that even in this condition the JSI is so thin and symmetric that the sample cannot generate a measurable modification in the HOM dip visibility. Note also that the photon wave packets remain identical, as in the case of free space propagation (Figure \ref{JSI_HOM_FILTER}$B$). Therefore, these simulations suggest that transmittance experiments under narrow bandwidth pump hardly would show changes in the HOM dip visibility due to ETPA.

\begin{figure*}[!h]
	\centering
	\includegraphics[width=\textwidth]{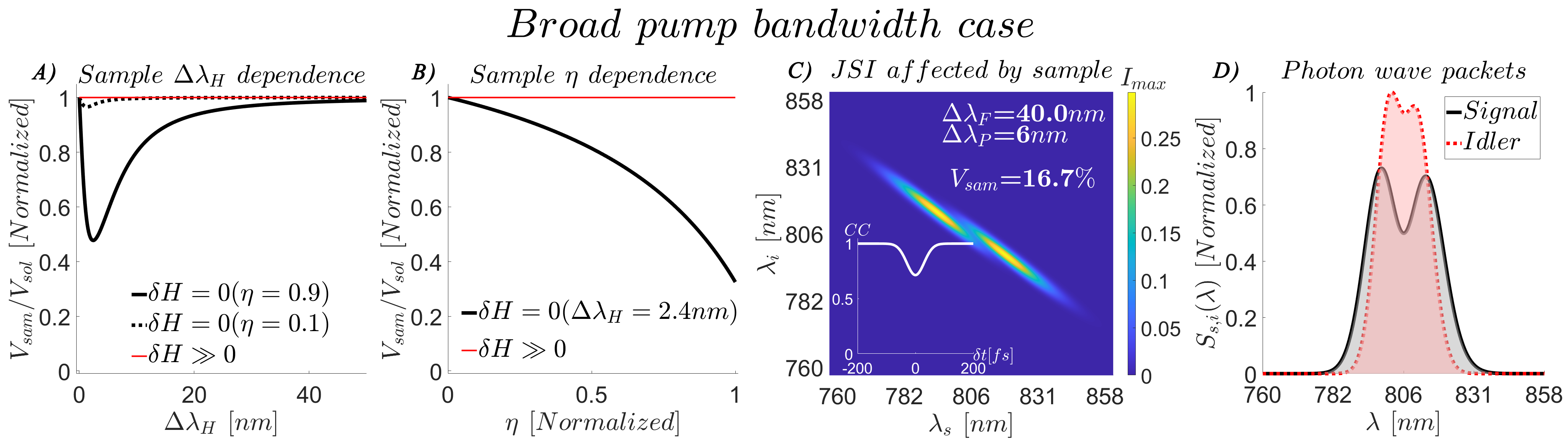}
	\caption{ETPA resulting from the sample interaction with entangled photons of broad bandwidth. $A$) $V_{sam}/V_{sol}$ as a function of sample bandwidth, $B$) $V_{sam}/V_{sol}$ as a function of the sample $\eta$ parameter, $C$) JSI affected by sample, $D$) single photon wave-packets.} 
	\label{PULSED_PUMP_CASE}
\end{figure*}

Figure \ref{PULSED_PUMP_CASE} presents the computations performed now assuming a pump of broad $\Delta\lambda_{p}$. All figure legend conventions remain the same as in the Figure \ref{CW_PUMP_CASE}. Contrary to the narrow bandwidth pump, we can see larger changes in $V_{sam}/V_{sol}$. However, the changes are still small when the nonlinearity of the sample is weak ($\eta=0.1$) as shown in Figure \ref{PULSED_PUMP_CASE}$A$ even under resonance ($\delta H=0$). In the extreme of an ideal sample in resonance bearing large nonlinearity ($\eta=0.9$), the same figure shows that $V_{sam}/V_{sol}$ would reach its minimum value at $\Delta\lambda_{H}\simeq 2.4nm$; such a sample bandwidth establishes an optimum sample characteristic to have the maximum change in visibility. In this case, according to Eq. \ref{RateETPA}, $58\%$ of the photons hitting the sample are absorbed by ETPA. By keeping an optimal sample bandwidth ($\Delta\lambda_{H}\simeq 2.4nm$) and resonance, the model predicts well pronounced changes for $V_{sam}/V_{sol}$ value as a function of $\eta$ as shown in Figure \ref{PULSED_PUMP_CASE}$B$, which means that such case the sample can produce a measurable ETPA. This can be seen more clearly in the modification of the JSI shown in Figure \ref{PULSED_PUMP_CASE}$C$, as well as the marked differences between the single photon wave-packets (Figure \ref{PULSED_PUMP_CASE}$D$).

As a brief summary of the results from the proposed model, in HOM interference two general facts have to be considered for ETPA experimental implementations: i) the use of broadband excitation, because it favors asymmetry in the JSI and consequently tend to generate differences in the photon wave-packets; ii) samples with small $\Delta\lambda_{H}$ and large $\eta$. These considerations are completely in agreement with previous works\cite{Dayan_exp,Dayan_teo}. 

\section{\label{sec:Experimental-section}Experiment}
\begin{figure*}[h]
	\centering
	\includegraphics[width=\textwidth]{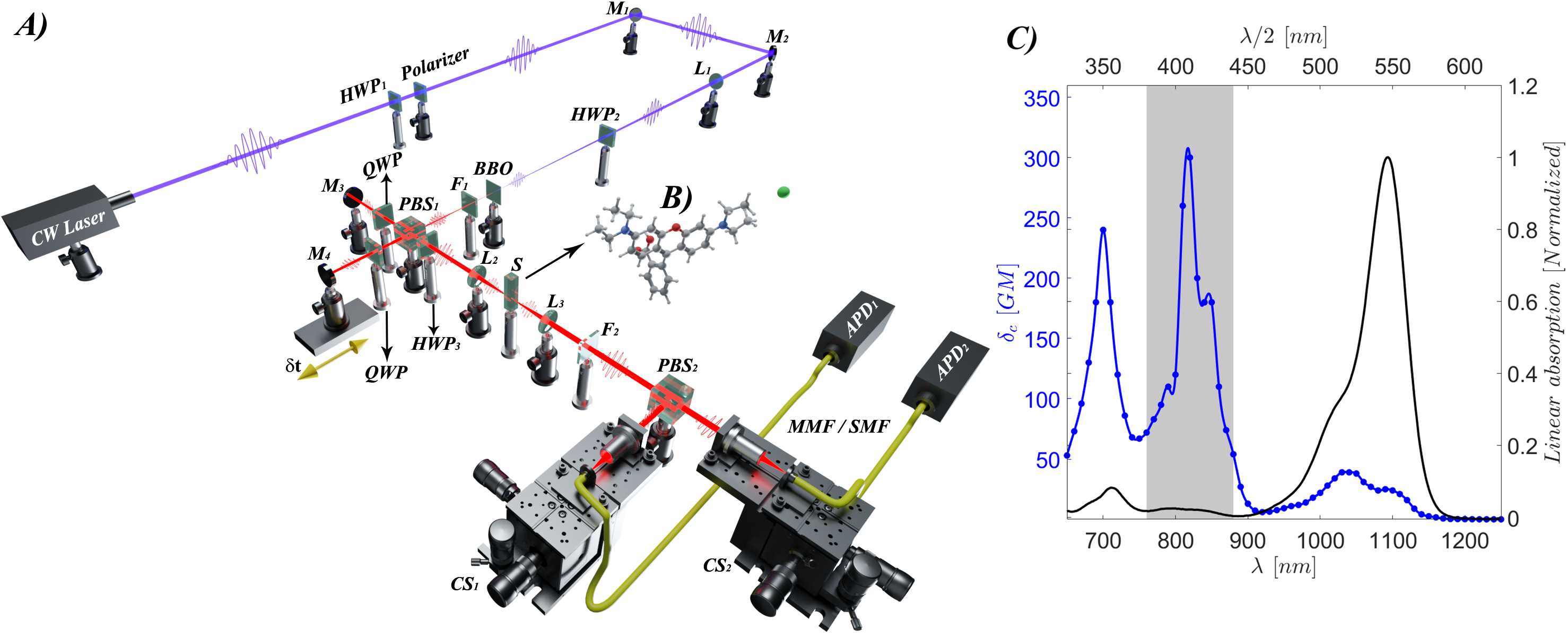}
	\caption{A) Experimental configuration for the transmittance ETPA experiments designed to use the HOM interferometer as a sensing device. MMF/SMF denote multi-mode and single-mode fiber, respectively. B) Rhodamine B molecule. C) Rhodamine B TPA cross-section (blue dots) and Rhodamine B linear absorption spectrum (black solid curve) as a function of wavelength. The gray rectangle corresponds to the the excitation region in our ETPA transmittance experiment. The experimental data represented as blue dots was taken from \cite{Makarov08}.} 
	\label{setup2}
\end{figure*}

\begin{figure}[h] 
	\centering
	\includegraphics[width=85mm,height=82mm]{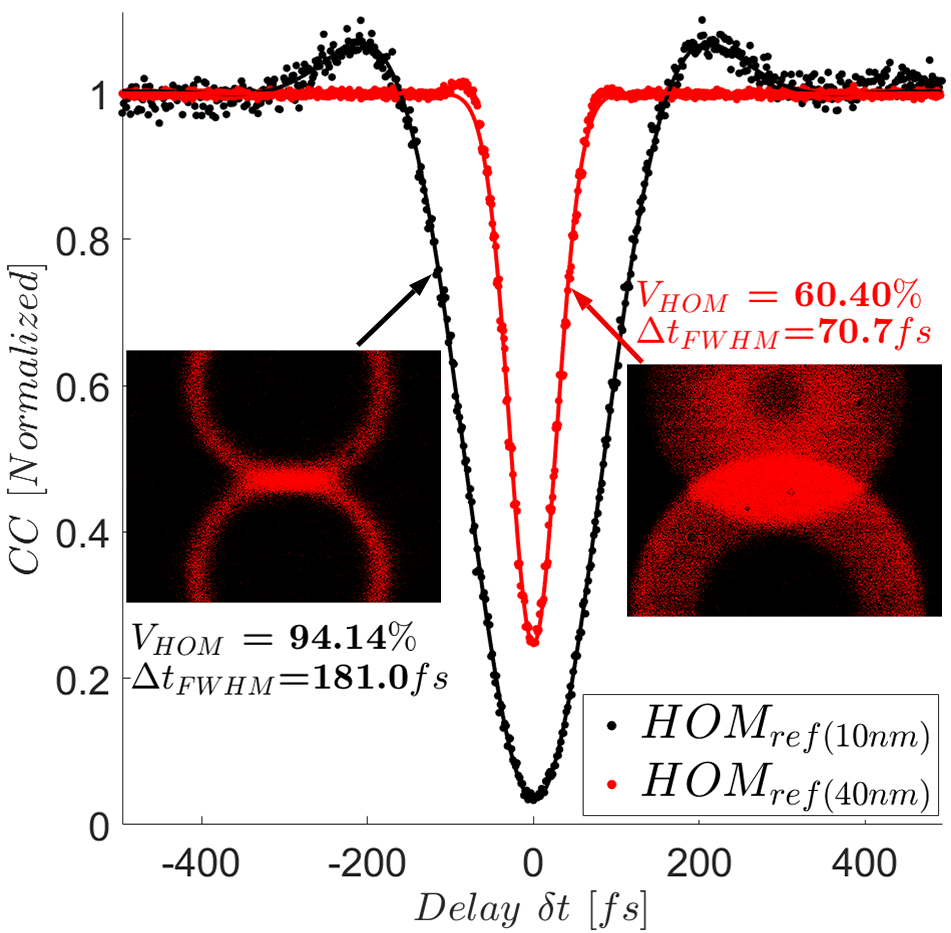}
	\caption{HOM dip obtained for free space propagation of the down-converted photons (cuvette removed), for two filter configurations. The black dots correspond to the HOM dip obtained with a $10$nm bandpass filter and longpass filter, while the red dots where obtained with a $40$nm bandpass filter and longpass filter. The insets show an image of the Type-II SPDC rings distribution as taken by a CCD camera (Thorlabs DCU224M), for both filter configurations. The collinear photons used in the experiments were obtained from the region where the SPDC cones overlap.} 
	\label{HOM_interferogram_calibration}
\end{figure}

The experimental setup is shown in Figure \ref{setup2}$A$. A CW laser (Crystalaser DL-405-100), centered at $\lambda_{p}=403nm$ and with a $\Delta\lambda_{p}\sim 1nm$ bandwidth FWHM, is focused by lens $L_{1}$ (focal length $f_1=500mm$) into a BBO crystal to produce collinear cross-polarized frequency-degenerate Type-II SPDC photons pairs around $806nm$. In order to optimize the phase-matching conditions, which maximize the SPDC emission process, the polarization of the pump beam is aligned with the plane defined by the crystal's optical axis. We do so by rotating the zero order half wave plate $HWP_{2}$, which also works as a control of the density of down-converted photons produced by the crystal. Prior being focused to enter the crystal, the laser pump power is controlled by a half-plate-wave ($HWP_{1}$) and a Glan-Thompson polarizer ($Polarizer$). The residual pump after the crytsal is eliminated by using a filter element $F_{1}$, composed of a longpass filter (Thorlbas FELH0500) and a bandpass filter (Thorlbas FBH800-40). To make sure that the residual pump was eliminated with $F_{1}$, a picture of the entangled-photons spatial mode was taken in two configurations of photon density obtained by varying $HWP_{2}$: 1) with maximum photon density and 2) minimum photon density. In the first case, a spatial mode like the one shown on the right side of Figure \ref{HOM_interferogram_calibration} is obtained, while in the second case no spatial mode is detected and the image is essentially dark, which indicates that there is no residual presence of pumping, more details can be found in the Supporting Information, Fig S3. Then, the filtered down-converted photons propagates through a Michelson interferometer (starting at $PBS_{1}$) which introduces a controllable temporal delay ($\delta t$) between them. 
In addition to the control of frequency ($F_{1}$) and temporal delay ($\delta t$) indistinguishability, the final requisite to obtain an optimal HOM interference pattern at $PBS_{2}$ is to eliminate the polarization distiguishability that is characteristic in photon pairs produced by Type-II SPDC process. To do so, a half-waveplate ($HWP_{3}$) is introduced to rotate $45^{\circ}$ the horizontal and vertical axis of polarization of the down-converted photons. This element works as a control to turn ``on'' or ``off'' the HOM interference. Then, the down-converted photons are focused into the $1cm$ quartz-cuvette containing the sample ($S$) with a $5cm$ focal length lens ($L_{2}$), producing a $W_{0}=58\mu m$ spot diameter which is then collimated with a second lens of the same focal length ($L_{3}$). A Rayleigh length of $Z_{R}=1.3cm$ is generated, so the interaction volume of the photons with the sample can be considered as a cylinder of $l=1cm$ length \cite{KristenM}.
	
The HOM effect results from the interference of indistinguishable photons at a beam splitter $PBS_{2}$ \cite{2PI_Fearn,2PI_Pittman,2PI_Legero,SPDC_Shih,Hong_Ou_Mandel1}. In our setup we obtain the HOM interferogram by recording the $CC$ as a function of the time delay between the photons, registered by a time-to-digital converter module (ID Quantique id800) and the avalance photodiodes $APD_{1}$ and $APD_{2}$ (Excelitas SPCM-AQRH). As a crucial property of the ETPA transmittance experiments discussed below, in our setup we can change the the degree of photons indistinguishability in all the relevant degrees of freedom (frequency, polarization, arrival time), allowing to fully control the interference process. A detailed experimental analysis of the interference process under different conditions of photons indistinguishability, noise and errors are found in the Supporting Information, Fig S4.

As an initial calibration step, we measured the HOM dip produced by  the down-converted photons in free space propagation ($HOM_{ref}$). As shown in Figure \ref{HOM_interferogram_calibration}, the distinctive characteristic of the $HOM_{ref}$ is a steep dip around $\delta t = 0$ and a high level of $CC_{max}$ for $\delta t$ values larger that the coherence length of the down-converted photons. In this figure two different configurations of photon indistinguishability are presented. The first configuration was set to optimize the spectral indistinguishability of the photon pairs by using a bandpass filter (Thorlbas FBH810-10) of narrow bandwidth ($10$nm, centered at $810$nm) and longpass filter (Thorlbas FELH0500) in the element $F_{1}$ in combination with a single-mode fiber (SMF), obtaining a $V_{ref}$ of $94\%$ and a temporal FWHM of $181$fs. For the second configuration a bandpass filter of $40$nm bandwidth centered at $800$nm (Thorlabs FBH800-40) and longpass filter (Thorlbas FELH0500) was used in combination with a multi-mode fiber (MMF), obtaining a $V_{ref}$=$60$\% and a FWHM temporal width of $71$fs. The reduction in $V_{ref}$ is due to asymmetric spectrum of each photon proper of a SPDC Type-II process \cite{SPDC_Grice}. Hereafter, in all the experiments the second filter configuration is employed.

\subsection*{\label{sec:Sample-solvent-configuration}Sample and solvent configuration}
The sample used as a model in our ETPA experiments to generate $HOM_{sam}$ was Rhodamine B ($\geq95\%$ purity, Sigma-Aldrich) $(RhB)$ dissolved in methanol at different concentrations: $0.1\mu$M, $1\mu$M, $0.01$mM, $0.1$mM, $1$mM, $10$mM, $58$mM, $100$mM. The well-known linear absorption spectrum of this molecule is displayed in Figure \ref{setup2}$C$ along with the nonlinear absorption spectrum obtained from the classical TPA effect \cite{Makarov08}. As it can be seen, around $800$nm there is region of two-photon resonance (indicated in gray color in the figure), which we will aim at in our experiments, unlike of previous work where the ETPA in $RhB$ was associated to the excitation of the state $S_{2}$ corresponding to a one-photon energy of $355$nm \cite{JuanVillabona1}. In order to avoid alignment errors, the cuvette containing the samples was fixed all the time. A special cleaning and drying process of the fixed cuvette was implemented every time we changed the sample under study. 

\section{\label{sec:Results}Experimental results and Discussion}
\begin{figure*}[!h]
	\centering
	\includegraphics[width=\textwidth]{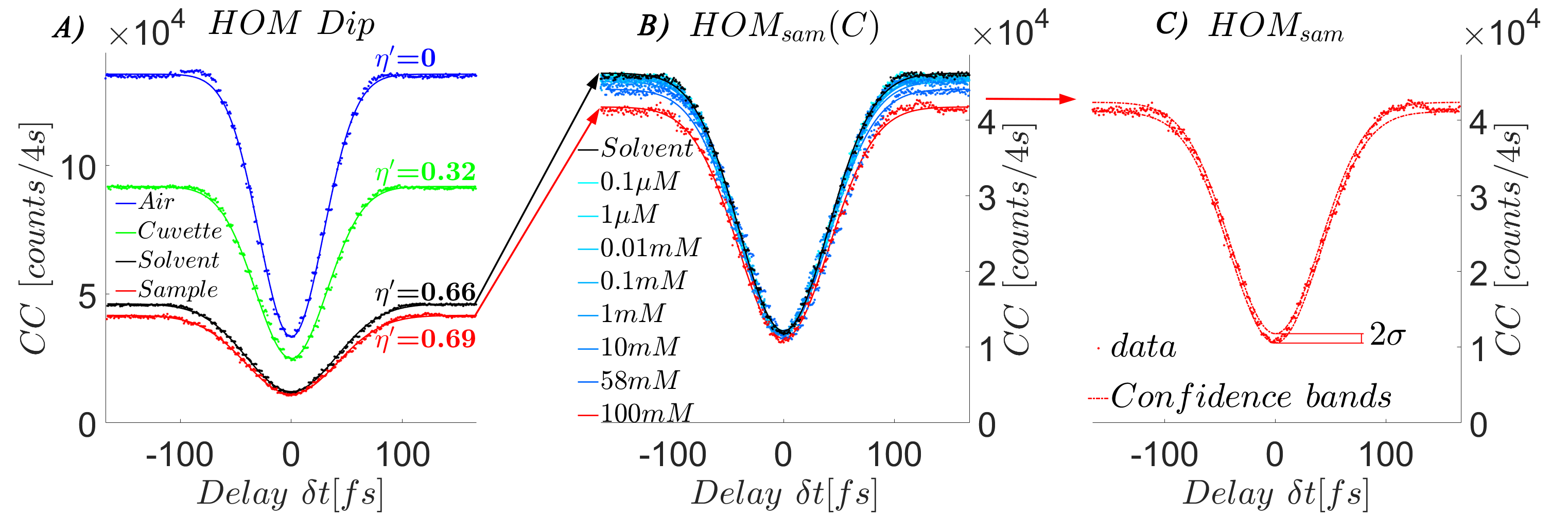}
	\caption{HOM dip results for $A$) air ($V_{ref}=60.4\pm 1.3\%$), cuvette ($V_{cuv}=60.2\pm 1.3\%$), solvent ($V_{sol}=58.3\pm 1.2\%$), and most concentrated sample ($V_{sam}=57.8\pm 1.4\%$), $B$) the set of concentrations of Rhodamine B in methanol, and $C$) the most concentrated sample with the calculated confidence bounds (fluctuation experimental data range). In $A$ and $B$, the continuous lines are the theoretical simulations from the model. } 
	\label{HOM_EXP}
\end{figure*}

In the experiments, the laser pump power was fixed at $43.9$mW. Figure \ref{HOM_EXP}$A$ presents four HOM dips of interest: $HOM_{ref}$ (free space propagation), $HOM_{sol}$ (solvent into the cuvette) and $HOM_{sam}$ (sample at the highest concentration); in addition, the case of a empty cuvette  (without solvent) is also shown.  
This figure shows that $CC_{max}$ decrease as the cuvette, solvent or samples are introduced in the path of propagating photons. As it can be clearly seen, well defined offsets are detected with respect to $HOM_{ref}$. The offset is large between $HOM_{ref}$ and $HOM_{sol}$ because the focused beam in the latter case travels through a denser medium changing the conditions of light coupling to the detection devices (APDs); meanwhile, the offset between $HOM_{sam}$ and $HOM_{sol}$ is small. This small offset between solvent and sample, clearly detectable at long $\delta t$, are ascribed to Fresnel losses, scattering and residual linear absorption, which are larger in the sample than in solvent. It is worth to remark that in our model the linear losses do not change the visibility of an interferogram. To see this, Figure S5 presents the simulation of a HOM dip obtained for a sample with different levels of lineal optical losses: the visibility remains unaltered when the losses varied in the range $0\leq \eta'< 1$. This outcome was expected as the lineal losses are independent of frequency, and this is exactly the experimental result observed for the HOM dips of Figure \ref{HOM_EXP}$A$, where all HOM dips have nearly the same visibility. Since $\eta'$  accounts in the model for linear losses unable to change the visibility while $\eta$ is related to the ETPA effect that inherently would change the visibility, then the experimental results suggest that the sample is not producing non-linear absorption.  Notice that the experimental HOM dips are reproduced accurately by the model (Eq. S1) considering the nonlinearity of the sample negligible ($\eta=0$) and different values of $\eta'$, as depicted by the continuous lines in Figure \ref{HOM_EXP}$A$ and Figure \ref{HOM_EXP}$B$. The $\eta'$ values (0 for free space, 0.32 for cuvette, 0.66 for solvent and 0.69 for most concentrated sample) generated by the model are in good agreement with the level of linear losses expected for each case. 

In Figure \ref{HOM_EXP}$B$ a zoom of the $HOM_{sol}$ and $HOM_{sam}$ dips for the entire sample concentrations set is displayed, where the black dots represent data from the solvent and red dots from the sample with the highest concentration (100mM); cyan dots are data for the other utilized concentrations in the range $0.1\mu$M - $58$mM. It is interesting to note that as the concentration increases, the offset in $CC$ is greater outside the dip (large delays) than inside the dip (zero delay), which is an intrinsic effect of the HOM dip as it was recently discussed\cite{Samuel_Alfred}. This intriguing effect depicted markedly in Figure \ref{HOM_EXP}$B$, apparently implies that the HOM visibilities decrease from solvent to samples of higher concentrations, however, by using Eq. \ref{VIS} it is found that  $V_{ref}$, $V_{sol}$, and all $V_{sam}$ are essentially the same within the range of uncertainties of the experimental error, the latter determined by the noise fluctuations of the $CC$ signal. To visualize the level of  experimental uncertainties, Figure \ref{HOM_EXP}$C$ presents as an example the data corresponding to the sample with the highest concentration; here the $HOM_{sam}$ is plotted within a known range of confidence bands, whose width represents 2 standard deviations (2$\sigma$) of the experimental data. Based in the 2$\sigma$ and the propagation of uncertainties when Eq. \ref{VIS} is employed, the visibility of each experimental HOM dip and its corresponding uncertainty was calculated. For instance, the visibility of the HOM dip for the sample at the heights concentration resulted in  $V_{sam}=57.8\pm 1.4\%$ while for the air  and solvent was $V_{ref}=60.4\pm 1.3\%$ and $V_{sol}=58.3\pm 1.2\%$. Table \ref{TableVIS} summarizes the visibilities obtained for each sample and the values obtained for the ratio $V_{sam}/V_{sol}$.  

\begin{figure*}[!h]
	\centering
	\includegraphics[width=\textwidth]{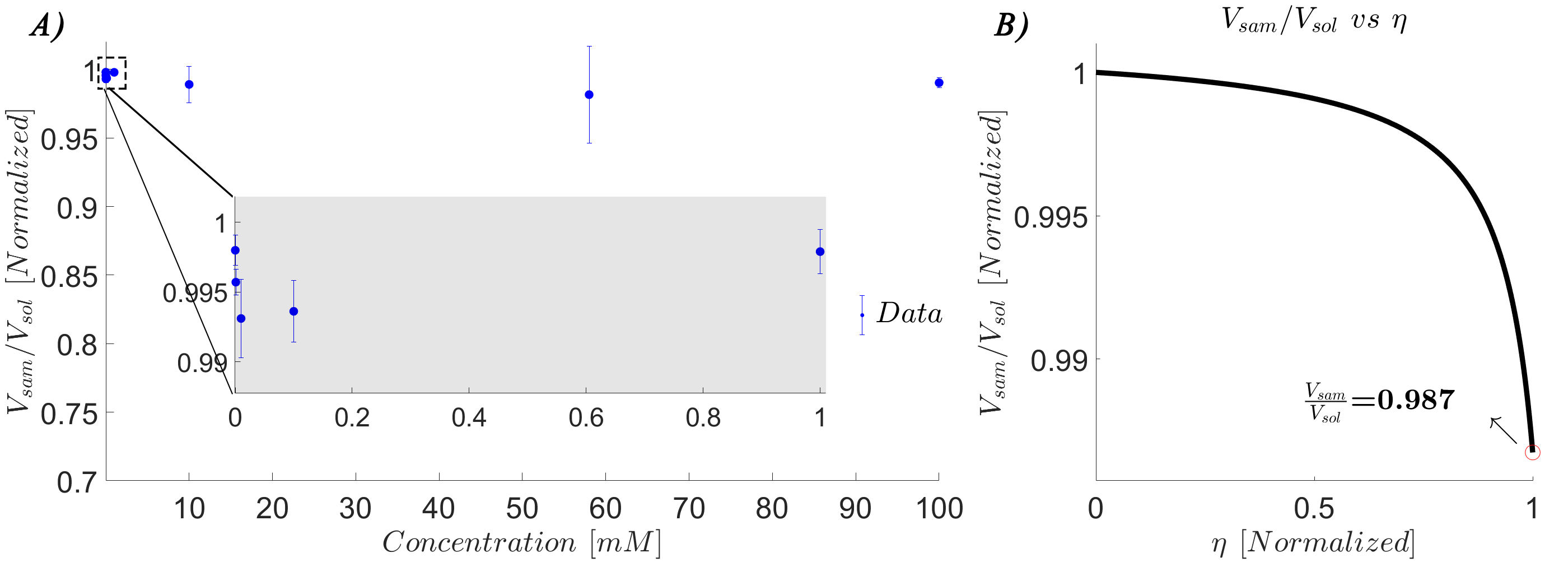}
	\caption{Experimental $V_{sam}/V_{sol}$ dependence with the sample concentration (panel A) and simulation of $V_{sam}/V_{sol}$ (panel B) as a function of $\eta$ highlighting, by means of red circle, the sample as a perfect ETPA absorber ($\eta=1$). In panel A blue dots are experimental data, and the gray area is the uncertainty range of the data.} 
	\label{VIS_CC_Concentration}
\end{figure*}

Figure \ref{VIS_CC_Concentration} shows the $V_{sam}/V_{sol}$ ratio as a function of the sample concentration, showing just marginal  changes in the visibility of samples respect to the solvent; the absence of a well defined trend in the variation of $V_{sam}/V_{sol}$ avoids to conclude unambiguously the presence of ETPA activity, at least with the sensitivity of our experimental apparatus. 
Using the mathematical model here introduced, we computed the changes of visibility predicted under our experimental conditions: sample bandwidth $\Delta\lambda_{H}=30nm$ (FWHM), central wavelength of the sample transfer function $2\lambda_{H}=816nm$ (taken directly from the TPA spectrum of Rhodamine B), pump bandwidth $\Delta\lambda_{p}\sim 1nm$ (FWHM) centered at $\lambda_{p}=403nm$, and filter bandwidth $\Delta\lambda_{F}=40nm$ (FWHM) centered at $\lambda_{F}=800nm$. After making a run of the model, no significant changes in visibility are found; only considering the sample as a perfect ETPA absorber (Figure \ref{VIS_CC_Concentration}$B$ for $\eta=1$) produces a $V_{sam}/V_{sol}=0.987$ value, which is just in the border of the experimental uncertainties of our apparatus. When a broadband pump is assumed, the ratio $V_{sam}/V_{sol}=0.935$ is obtained; such a change in the visibility is in principle detectable in our experimental approach, but it implies the unrealistic condition of a perfect ETPA absorber.  
\begin{table}[]
\centering
\begin{tabular}{|c|c|c|c|c|}
\hline
$C[M/L]$               & $V_{sam}[\%]$       & $V_{sam}/V_{sol}$                                  \\ \hline
$1\times10^{-7}$                & 58.2$\pm$1.2                                  & 0.998$\pm$0.001                                 \\
$1\times10^{-6}$                & 58.1$\pm$1.2                                  & 0.996$\pm$0.001                                 \\
$1\times10^{-5}$                & 57.9$\pm$1.3                                  & 0.993$\pm$0.003                                    \\
$1\times10^{-4}$               & 58$\pm$1.3                                  & 0.994$\pm$0.002                                  \\
$1\times10^{-3}$               & 58.2$\pm$1.2                                  & 0.998$\pm$0.002                                  \\
$1\times10^{-2}$                & 57.7$\pm$1.9                                  & 0.99$\pm$0.01                                  \\
$1\times10^{-2}$               & 57.3$\pm$3.2                                  & 0.98$\pm$0.04                                  \\
$1\times10^{-1}$                & 57.8$\pm$1.4                                  & 0.990$\pm$0.004                                  \\ \hline
\end{tabular}
\caption{Experimental values of the visibilities for the HOM dips obtained from samples at different molar concentration ($C$); change of the sample visibility respect to the solvent visibility and uncertainty error range of the data. }
\label{TableVIS}
\end{table}

From the mathematical and experimental considerations above discussed, it is difficult and probably unworkable to detect ETPA in Rhodamine B in the $800nm$ region, because the effect strongly depends on the detuning between the sample transfer function and  the filtered photons function. Physically, this means that as the spectral sample bandwidth differs markedly from the corresponding spectrum of  photons, less is the probability of inducing JSI asymmetries susceptible of being detected as changes in the HOM dip. When the spectral considerations are not fulfilled, the ETPA interaction can not be reached, so the sample behaves as a photon attenuator which eliminates equally photons of all frequencies (a neutral density filter characterized by $\eta'$), producing an offset between $HOM_{sol}$ and $HOM_{sam}$.

This study suggests that previous works \cite{JuanVillabona1} in which Rhodamine B was used as a model to study ETPA might involve signals not coming from two-photon absorption, but from artifacts or spurious signals, i.e., linear losses. Transmittance experiments intended to measure $R_{TPA}$ based exclusively on photon counting have  procedural and theoretical problems to discriminate optical losses other than TPA. In this context, a system based on the interference of two photons, in which visibility of a HOM dip is not affected by linear optical losses, represents a novel and alternative scheme to detect the changes in the photon symmetry of the correlated photon-pairs state produced by the non-linear absorption effect. 

\section{Conclusions}
In this paper we have used the two-photon interference effect as a mechanism to study and establish the spectral considerations for which ETPA transmittance experiments can be carried out, analyzing the visibility of the HOM dip, being the first time, to the best of our knowledge, that this approach is explored.  

We presented a mathematical model for the interaction of  down-converted photons with a nonlinear sample. The model uses the parameter $\eta$ to account for the  strength of the ETPA process. In order predict the spectral conditions that would lead to measurable ETPA by means of our experimental proposal, theoretical simulations were generated using as variables the sample bandwidth and the detuning of the photon spectrum, under either narrow or broad bandwidth pump. By considering the visibility of the HOM dips as figure of merit, we did not find experimental evidence of ETPA, in agreement with the simulations that compare the visibilities a solution of Rhodamine B and the solvent alone. Thus, it can be inferred that the significant detuning between the sample spectrum and the photon pair spectrum, in combination with possibly weak sample nonlinearities, precluded introducing asymmetries in the JSI, a necessary condition to induce changes in the visibility of HOM dips. 

Despite the ETPA was not unequivocally determined in our experiments for the case of Rhodamine B, our model allowed to explore the  combinations of pumping bandwidth, detuning, sample bandwidth and sample nonlinear strength that would favor changes in the visibility of HOM dips, and in turn the detection of ETPA signals under a transmission scheme. For this scheme, the effect of linear losses independent of frequency was also explored. Considering this, we believe that our experimental and theoretical results represent a step forward in the application of quantum sensing techniques as ultra-sensitive devices for the study of elusive nonlinear optical effects, such as ETPA in molecular systems.

\section{Data Availability Statement}
        Data underlying the results presented in this paper are not publicly available at this time but may be obtained from the authors upon reasonable request.
        
\section{Acknowledgments}
We acknowledge support from CONACYT, Mexico. This work was supported by CONACYT, Mexico grant FORDECYT-PRONACES 217559.

\section{Disclosures}
The authors declare no conflicts of interest.


\end{document}